\documentclass[journal,10pt,twocolumn]{IEEEtran}
\usepackage{amssymb,relsize,graphicx,amsmath,psfig,cite,float,subfigure,setspace,hyperref,flushend}
\usepackage{xcolor}
\usepackage{tabularx}
\usepackage{algorithm}
\usepackage{algpseudocode}

\newtheorem{prop}{Proposition}

\newcommand{\bi}{\begin{itemize}}
\newcommand{\ei}{\end{itemize}}

\newcommand{\be}{\begin{enumerate}}
\newcommand{\ee}{\end{enumerate}}
\newcommand{\bd}{\begin{description}}
\newcommand{\ed}{\end{description}}
\newcommand{\bc}{\begin{center}}
\newcommand{\ec}{\end{center}}
\newcommand{\bt}{\begin{tabbing}}
\newcommand{\et}{\end{tabbing}}
\newcommand{\bfig}{\begin{figure}}
\newcommand{\efig}{\end{figure}}
\newcommand{\beq}{\begin{equation}}
\newcommand{\beqarr}{\begin{eqnarray}}
\newcommand{\beqarrn}{\begin{eqnarray*}}
\newcommand{\eeq}{\end{equation}}
\newcommand{\eeqarr}{\end{eqnarray}}
\newcommand{\eeqarrn}{\end{eqnarray*}}
\newcommand{\bflr}{\begin{flushright}\vspace{-0.2in}}
\newcommand{\eflr}{\end{flushright}}
\newcommand{\bsub}{\begin{subequations}}
\newcommand{\esub}{\end{subequations}}
\newcommand{\barr}{\begin{array}}
\newcommand{\earr}{\end{array}}
\newcommand{\nn}{\nonumber}

\def\undb#1{\mbox{\bf{#1}}}

\def\dn{\stackrel{\scriptscriptstyle \triangle}{=}}

\def\arg{\mbox{arg}}

\begin{document}
\bstctlcite{bstctl:nodash}

\title{\huge{RIS-Assisted Noncoherent Wireless System: Error Analysis with Optimal Receiver and Multi-level ASK}}

\author{Sambit Mishra, \IEEEmembership{Student Member, IEEE} and Soumya P. Dash, \IEEEmembership{Senior Member, IEEE}
\thanks{The authors are with the School of Electrical and Computer Sciences, Indian Institute of Technology Bhubaneswar, Argul, Khordha, India, email: 21ec01021@iitbbs.ac.in,  soumyapdashiitbbs@gmail.com.}
}

{}

\maketitle

\begin{abstract}
This paper considers a reconfigurable intelligent surface (RIS) aided wireless communication system where the transmitter employs one-sided amplitude-shift keying (ASK) for data modulation and the receiver employs an optimal noncoherent maximum-likelihood rule for symbol detection. A novel statistical analysis is presented to approximate the weighted sum of a central and a non-central chi-squared random variable, which is used to derive a novel closed-form expression for the symbol error probability (SEP) of the noncoherent system. Furthermore, an optimization problem to minimize the system's SEP under transmission energy constraint is proposed, and novel algorithms are presented to obtain the optimal ASK constellation minimizing the SEP. Numerical results indicate that the noncoherent system achieves superior error performance and higher diversity order with the optimal ASK constellation compared to the traditional equispaced ASK constellation. Further, the optimal ASK significantly differs from the traditional one, with the difference becoming significant with an increase in the average signal-to-noise ratio and at a lower number of RIS's reflecting elements.
\end{abstract}

\begin{IEEEkeywords}
Amplitude-shift keying; Noncoherent communication; Optimization; Reconfigurable intelligent surfaces; Symbol error probability.
\end{IEEEkeywords}

\IEEEpeerreviewmaketitle

\section{Introduction}

Sixth-generation (6G) wireless networks envision transformative services such as holographic communication, extended reality, tactile Internet, and global connectivity \cite{JiHa21}. These applications demand a significant increase in network capacity and efficiency, necessitating advancements in energy-efficient architectures and reduced hardware complexity \cite{WaYo23}. While technologies like massive multiple-input multiple-output (MIMO), millimeter wave communication, and ultra-dense networks have partially addressed 5G limitations, low energy and hardware demands still remain challenging aspects.

In this context, reconfigurable intelligent surface (RIS) and noncoherent communications have emerged as promising technologies for 6G \cite{TaCh21, 10041914,10385147,10040911}. RIS consists of passive, low-cost metasurfaces composed of reconfigurable elements that manipulate electromagnetic waves based on the channel conditions, thus enhancing signal coverage, reducing path loss, and improving interference mitigation without requiring active RF chains \cite{10753596, 10400440}. This has led to extensive research to apply RIS for improving energy and spectral efficiency \cite{TaPe23, WaQi24}, enhancing coverage in terrestrial and aerial networks \cite{MeZh23, 10463684}, and enabling advancements in free-space optical systems \cite{NaSc21}.
A majority of the studies of RIS-assisted wireless systems assume the availability of the channel state information (CSI) at the transceiver pair, aiding their optimal design for performance maximization. Although acquiring CSI at the transmitter end is feasible owing to the available computational resources at the base station, the estimation of the instantaneous CSI puts a toll on the receiver, especially for hardware- and energy-constrained applications. The design of noncoherent receivers addresses this concern, thus leading to recent studies on RIS-assisted noncoherent wireless systems \cite{Se22, CaHuBa23, CaHuAn23, InWa24,  mishra2024optimal, mukhopadhyay2024optimal, 10547354}.

Motivated by this, we study a noncoherent wireless system in this paper, where the communication between the transceiver pair is supported by a RIS consisting of $L$ reflecting elements. Owing to the advantages of its generation and reception, the transmitter employs a multi-level one-sided amplitude-shift keying (ASK) modulation scheme, and the energy- and hardware-constrained receiver does not invest its resources in estimating the instantaneous CSI and utilizes a noncoherent structure for symbol detection. The key contributions of the paper are summarized as follows:
\begin{itemize}
\item Optimal maximum-likelihood (ML) receiver structure is derived for the scenario of the phase-shifts of the RIS being adjusted based on the channels in the transmitter-RIS and the RIS-receiver links.
\item A statistical analysis is proposed to approximate the weighted sum of a noncentral and a central chi-squared random variable, which is further used to derive the closed-form expression for the union bound on the symbol error probability (SEP) of the system.
\item An optimization framework is proposed to minimize the system's SEP under an average transmit energy constraint. Novel algorithms are presented to obtain optimal one-sided ASK constellations, achieving the objective of the optimization formulation.
\end{itemize}
Numerical analysis demonstrates that the statistical approximation leads to tight results in the SEP with the exact SEP values. Furthermore, the optimal ASKs are observed to outperform the traditional equispaced ones in terms of achieving lower SEP and higher diversity order values. 
\section{System Model}
We consider a system model where a transmitter, equipped with a single antenna, transmits multi-level ASK modulated symbols to a low-hardware-complex receiver. Additionally, the communication between the transceiver pair is assisted by a RIS consisting of $L$ reflecting elements. Considering that the receiver deploys symbol-by-symbol detection, the received signal is expressed as
\beq
r = \left( \undb{h}_2^H \mathbf{\Phi} \undb{h}_1 + h_d \right) s + n \, ,
\label{eq1}
\eeq
where $s$ is the transmitted symbol, $\undb{h}_1$ and $\undb{h}_2$ are the $L \times 1$ channel gain vectors modeling the transmitter-RIS and the RIS-receiver links, respectively, $h_d$ is the direct channel between the transceiver pair, $\left( \cdot \right)^H$ denotes the Hermitian (complex conjugate) operator, $\mathbf{\Phi}$ is a $L \times L$ diagonal matrix with its diagonal entries comprising of the phase shifts introduced by the reflecting elements of the RIS, and $n$ is the additive noise following a zero-mean complex Gaussian distribution, implying that $n \sim {\mathcal{CN}} \left(0,\sigma_n^2 \right)$.

Considering the fact that a RIS is typically deployed to enable communication channels with prominent line-of-sight (LoS) links between the transceiver pair, when the direct channel between them lacks such links, we statistically model the channel gains between the transmitter-RIS and the RIS-receiver pairs to follow complex Gaussian distributions with non-zero means, implying that $\undb{h}_1 \sim {\mathcal{CN}} \left( \mu_1 \undb{1}_L , \sigma_h^2 \undb{I}_L \right)$ and $\undb{h}_2 \sim {\mathcal{CN}}\left( \mu_2 \undb{1}_L , \sigma_h^2 \undb{I}_L \right)$, where $\undb{1}_L$ and $\undb{I}_L$ denote the $L \times 1$ vector of ones and $L \times L$ identity matrix, respectively, and, the direct channel $h_d$ to follow a zero-mean complex Gaussian distribution implying that $h_d \sim {\mathcal{CN}} \left(0,\sigma_{h_d}^2 \right)$. Furthermore, we consider that the transmitter has sufficient hardware complexity to estimate the phases of the compounded channel $\undb{h}_2^H \undb{h}_1$ and the phase shifts of the RIS elements are set accordingly to maximize the instantaneous power of the signal component arising due to the compounded channel. Thus, if we denote $h_{1,\ell} = \left| h_{1,\ell} \right| e^{\jmath \angle h_{1,\ell}}$ and $h_{2,\ell} = \left| h_{2,\ell} \right| e^{\jmath \angle h_{2,\ell}}$, with $\jmath=\sqrt{-1}$, then the corresponding $\ell$-th diagonal element of the phase matrix is $\left( \mathbf{\Phi} \right)_{\ell,\ell} = e^{\jmath \left(\angle h_{2,\ell} - \angle h_{1,\ell} \right)}$ for $\ell=1,\ldots,L$. This results in the received symbol in (\ref{eq1}) to be given as
\beq
r = \left( \sum_{\ell=1}^L \left| h_{1,\ell} \right| \left| h_{2,\ell} \right| + h_d \right) s + n \, .
\label{eq2}
\eeq
As a RIS typically consists of a large number of reflecting elements, we use the central limit theorem (CLT) to model $\sum_{\ell=1}^L \left| h_{1,\ell} \right| \left| h_{2,\ell} \right|$ as a Gaussian random variable as
\beqarr
\sum_{\ell=1}^L \left| h_{1,\ell} \right| \left| h_{2,\ell} \right|
\sim {\mathcal{N}} \left( \mu_f , \sigma_f^2 \right) \, ,
\label{eq3}
\eeqarr
where $\mu_f$ and $\sigma_f^2$ are obtained as
\bsub
\beqarr
\mu_f = \alpha \sigma_h^2 \, , \quad
\sigma_f^2 = \beta \sigma_h^4 \, ,
\label{eq4a}
\eeqarr
and $\alpha$ and $\beta$ are given by
\beqarr
\alpha \! \! \! \! &=& \! \! \! \! \frac{L \pi}{4}
L_{1/2} \left( - K_1 \right) L_{1/2} \left( - K_2 \right) \, , \nn \\
\beta \! \! \! \! &=& \! \! \! \! L 
\! \left[ \left(1+K_1 \right) \left(1+K_2 \right)
\! - \! \frac{\pi^2}{16} L_{1/2}^2 \! \left(-K_1 \right)
L_{1/2}^2 \! \left(-K_2 \right) \right] \, . \nn \\
\label{eq4b}
\eeqarr
\esub
Further, $K_1=\left| \mu_1 \right|^2/\sigma_h^2$ and $K_2=\left| \mu_2 \right|^2/\sigma_h^2$ denote the Rician parameters of the transmitter-RIS and RIS-receiver channels, respectively, and $L_{1/2} \left( \cdot \right)$ is a Laguerre polynomial.

The transmitted symbol $s$, chosen from the set of equiprobable multi-level ASK constellation, is expressed as $s_m = \sqrt{E_m}$, where $E_m , \, m=1,\ldots,M$, denotes the energy of the $m$-th symbol. Without loss of generality, we consider the symbols to be in ascending order, implying that $\sqrt{E_m} < \sqrt{E_{m+1}} \, , \forall m \in \left\{ 1, \ldots, M-1\right\}$.

The receiver structure, with the requirement of low hardware complexity, employs the optimal noncoherent ML detection rule as
\beq
\hat{s} = \arg \max_{s} f \left( r \big| s \right) \, ,
\label{eq5}
\eeq
where $f \left( r \big| s \right)$ is the conditional probability density function (p.d.f.) of the received signal $r$ conditioned on $s$. The computation of the p.d.f. in (\ref{eq5}) requires the statistics of the overall wireless channel, which, using (\ref{eq3}), is computed as
\beqarr
&& \Re \left\{ \sum_{\ell=1}^L
\left|h_{1,\ell} \right| \left|h_{2,\ell} \right| + h_d \right\}
\sim {\mathcal{N}} \left(\mu_f , \sigma_f^2 + \frac{\sigma_{h_d}^2}{2} \right) \, , \nn \\
&& \Im \left\{ \sum_{\ell=1}^L
\left|h_{1,\ell} \right| \left|h_{2,\ell} \right| + h_d \right\}
\sim {\mathcal{N}} \left( 0 , \frac{\sigma_{h_d}^2}{2} \right) \, .
\label{eq6}
\eeqarr
Thus, utilizing the statistics of the noise and the statistical independence of the real and imaginary parts of the received symbol, the optimal noncoherent ML receiver is obtained as
\beqarr
&&\!\!\!\!\!\!\!\!\!\!\!\!\!\!\!\!\!\!\!\!\!\!
\hat{s} = \arg \max_{s}
f \left( \Re \left\{ r \right\} \big| s \right) f \left( \Im \left\{ r \right\} \big| s \right)  \nn \\
&&\!\!\!\!\!\!\!\!\!\!\!\!\!\!\!\!\!\!
= \arg \min_{s}
\frac{\left( \Re \left\{r \right\} - \mu_f s \right)^2}
{\left(2 \sigma_f^2 + \sigma_{h_d}^2 \right) s^2 + \sigma_n^2}
+  \frac{\left( \Im \left\{r \right\} \right)^2} {\sigma_{h_d}^2 s^2 + \sigma_n^2}
\nn \\ 
&& \!\!\!\!\!\!\!
+ \frac{1}{2}
\ln \left( \left(2 \sigma_f^2 + \sigma_{h_d}^2 \right) s^2 \!+\! \sigma_n^2 \right)
+ \frac{1}{2}
\ln \left(  \sigma_{h_d}^2 s^2 \!+\! \sigma_n^2 \right).
\label{eq7}
\eeqarr
\section{Performance Analysis}
The performance of the receiver structure is evaluated in terms of the system's SEP, which is derived in this section. Owing to the structure of the receiver in (\ref{eq7}), we compute a bound on the SEP by using the union bound technique. Considering the transmission of the $m$-th symbol, i.e., $s_m=\sqrt{E_m}$, the pairwise probability of error (PEP) with a symbol $s_k$ is expressed as
\beqarr
&&\!\!\!\!\!\!\!\!\!\!\!\!\!
P_{e_{m \rightarrow k}}
= \Pr \left\{ \frac{\left( \Re \left\{r \right\} - \mu_f s_m \right)^2}
{\left (2 \sigma_f^2 + \sigma_{h_d}^2 \right) s_m^2 + \sigma_n^2} +  \frac{\left( \Im \left\{r \right\} \right)^2} {\sigma_{h_d}^2 s_m^2 + \sigma_n^2} \right. \nn \\
&& \quad \left. + \frac{1}{2} \ln \left( \left (2 \sigma_f^2 + \sigma_{h_d}^2 \right ) s_m^2 + \sigma_n^2 \right) + \frac{1}{2}
\ln \left(  \sigma_{h_d}^2 s_m^2 \!+\! \sigma_n^2 \right) \right. \nn \\
&& \quad \left. >  \frac{\left( \Re \left\{r \right\} - \mu_f s_k \right)^2} 
{\left (2 \sigma_f^2 + \sigma_{h_d}^2 \right ) s_k^2 + \sigma_n^2} + \frac{\left( \Im \left\{r \right\} \right)^2} {\sigma_{h_d}^2 s_k^2 + \sigma_n^2} 
\right. \nn\\
&& \quad
\left. + \frac{1}{2} \ln \left( \left (2 \sigma_f^2 + \sigma_{h_d}^2 \right ) s_k^2 + \sigma_n^2 \right)  \!+\! \frac{1}{2}
\ln \left(  \sigma_{h_d}^2 s_k^2 \!+\! \sigma_n^2 \right)\right\}\! \nn \\
&& \qquad \qquad \qquad \qquad \qquad \quad \,\,\,\,k=1,\ldots, M,\, k \neq m \, .
\label{eq8}
\eeqarr
We define the terms ${X}_m$, $Y_m$, $D_m$, and ${B}_m$ as
\beqarr
{X}_m \! \! \! \! & \dn & \! \! \! \!
\left( \sum_{\ell=1}^L
\left| h_{1,\ell} \right| \left| h_{2,\ell} \right|
+ \Re \left \{ h_d \right \} - \mu_f \right) \sqrt{E_m}
+ \Re \left\{ n \right\} \, , \nn \\
Y_m \! \! \! \! & \dn & \! \! \! \!
\left( \Im \left\{h_d \right\} \right) \sqrt{E_m}
+ \Im \left\{n \right\} \, , \nn \\
{B}_m \! \! \! \! & \dn & \! \! \! \!
\left (2 \sigma_f^2 + \sigma_{h_d}^2 \right ) E_m
+ \sigma_n^2 \, , \nn \\
D_m \! \! \! \! & \dn & \! \! \! \!
\sigma_{h_d}^2 E_m + \sigma_n^2,
\quad m= 1, \dots , M \, .
\label{eq9}
\eeqarr
We observe that for $E_m<E_k$, we have $B_k > B_m$ and $D_k > D_m$, and for $E_m>E_k$, we have $B_k < B_m$ and $D_k < D_m$. Thus, we evaluate the expression in (\ref{eq8}) for these two cases.
\underline{\textbf{For $m<k$}}, the PEP in (\ref{eq8}) can be rewritten as
\beqarr
&& \! \! \! \! \! \! \! \! \! \! \! \! \! \! \! \! \!
P_{e_{m \rightarrow k}}
= \Pr \left\{ \frac{X_m^2}{B_m}
- \frac{\left(X_m - \mu_f \left(\sqrt{E_k}
- \sqrt{E_m} \right) \right)^2}{B_k} \right. \nn \\
&& \left. + \frac{Y_m^2}{D_m} - \frac{Y_m^2}{D_k}
> \frac{1}{2}\ln\left(\frac{B_kD_k}{B_mD_m}\right) \right\} \nn \\
&& \! \! \! \! \! \! \! \! \! \! \! \! \! \! \!
= \Pr \left\{\left(B_m^{-1} - B_k^{-1}\right)
\left(X_m - \frac{\mu_f B_m \left(\sqrt{E_m} - \sqrt{E_k} \right)}
{\left(B_m - B_k \right)} \right)^2 \right. \nn \\
&& \quad
\left. + \left(D_m^{-1} - D_k^{-1}\right) Y_m^2 
> \frac{1}{2}\ln\left(\frac{B_k D_k}{B_m D_m}\right) \right. \nn \\
&& \qquad \qquad \qquad \qquad \quad
\left. + \frac{\mu_f^2\left( \sqrt{E_k} - \sqrt{E_m} \right)^2}
{\left(B_k - B_m\right)} \right\} .
\label{eq10}
\eeqarr
Similarly, \underline{\textbf{for $m>k$}}, the PEP in (\ref{eq8}) is simplified as
\beqarr
&& \! \! \! \! \! \! \! \! \! \! \! \! \! \! \!
P_{e_{m \rightarrow k}} \nn \\
&& \! \! \! \! \! \! \! \! \! \! \! \! \! \! \!
= \Pr \left\{ \left( B_k^{-1} - B_m^{-1} \right)
\left( X_m - \frac{\mu_f B_m \left(\sqrt{E_m} - \sqrt{E_k} \right)}
{\left(B_m - B_k\right)} \right)^2  \right. \nn \\
&& \quad \left. + \left(D_k^{-1} - D_m^{-1}\right) Y_m^2
< \frac{1}{2}\ln\left(\frac{B_mD_m}{B_kD_k}\right) \right. \nn \\
&& \qquad \qquad \qquad \qquad \quad
\left. + \frac{\mu_f^2\left(\sqrt{E_m} - \sqrt{E_k} \right)^2}
{\left(B_m - B_k\right)} \right\} .
\label{eq11}
\eeqarr

It is to be observed from (\ref{eq10}) and (\ref{eq11}) that the effective random variable on the left-hand side of these expressions is a weighted sum of a central and non-central random variable. Thus, to compute the probabilities in these expressions, we first need to compute the statistics of the weighted sum, which is derived in the following proposition.
\begin{prop}
Let $X$ and $Y$ be two real and independent Gaussian random variables with $X \sim \mathcal{N} \left( \mu_x, \sigma_x^2 \right)$ and $Y \sim \mathcal{N} \left( 0, \sigma_y^2 \right)$. Further, let us define a random variable $Z$ as
\beq
Z \dn \eta X^2 + \delta Y^2 \ , 
\label{eq12}
\eeq
Then $Z$ approximately follows a Gamma distribution as $Z \approx \hat{Z} \sim \Gamma \left( \zeta, \theta \right)$, where
\beqarr
\zeta \! \! \! \! &=& \! \! \! \!
\frac{\left(\eta\left(\sigma_x^2 + \mu_x^2\right)
+ \delta\sigma_y^2\right)^2}
{2\eta^2\sigma_x^2\left(\sigma_x^2 + 2\mu_x^2\right)
+ 2\delta^2\sigma_y^4 } \, , \nn \\
\theta \! \! \! \! &=& \! \! \! \! \frac{2\eta^2\sigma_x^2\left(\sigma_x^2 + 2\mu_x^2\right)
+ 2\delta^2\sigma_y^4 }{\eta\left(\sigma_x^2 + \mu_x^2\right)
+ \delta\sigma_y^2} \, .
\label{eq13}
\eeqarr
\end{prop}
\begin{IEEEproof}
If $Z \approx \hat{Z}$, we use the moment matching method to ensure that $\undb{E} \left[Z \right] = \undb{E} \left[ \hat{Z} \right]$ and $\undb{E} \left[\left(Z - \undb{E} \left[ Z \right] \right)^2 \right] = \undb{E} \left[ \left( \hat{Z} - \undb{E} \left[ \hat{Z} \right] \right)^2 \right]$, which from the statistics of $X$ and $Y$, and of a Gamma distributed random variable results in
\beqarr
\zeta\theta \! \! \! \! &=& \! \! \! \!
\eta\left(\sigma_x^2 + \mu_x^2\right) + \delta\sigma_y^2 , \nn \\
\zeta\theta^2 \! \! \! \! &=& \! \! \! \!
2\eta^2 \sigma_x^2 \left(\sigma_x^2 + 2\mu_x^2\right)
+ 2\delta^2\sigma_y^4 \, .
\label{eq14}
\eeqarr
Solving the expressions in (\ref{eq14}) leads to the expressions of $\zeta$ and $\theta$ in (\ref{eq13}). This completes the proof.
\end{IEEEproof}


Utilizing the result in {\em Proposition 1} followed by algebraic simplifications on (\ref{eq10}) and (\ref{eq11}) by using the cumulative distribution function of a Gamma random variable, we obtain the upper bound on the SEP of the system as
\beqarr
&& \! \! \! \! \!
\! \! \! \! \! \!
P_{e} = \frac{1}{M}\sum_{m=1}^M \left[ \sum_{k=1}^{m-1} P_{e_{m \rightarrow k}}
+ \! \! \! \!
\sum_{k=m+1}^{M} P_{e_{m \rightarrow k}} \right] \nn \\
&& \! \!
\leq \frac{\left(M^2-1\right)}{6}
- \frac{1}{M}\sum_{m=1}^{M} \left[ \sum_{k=1}^{m-1}
\frac{1}{\Gamma\left(\zeta_{m,k}\right)}
\gamma\left(\zeta_{m,k}, \frac{\mathcal{K}_{m,k}}
{\theta_{m,k}}\right) \right. \nn \\
&& \qquad \qquad \quad 
\left. - \sum_{k=m+1}^{M}
\frac{1}{\Gamma\left(\zeta_{m,k}\right)}
\gamma\left(\zeta_{m,k}, 
\frac{\mathcal{K}_{m,k}}{\theta_{m,k}}\right) \right],
\label{eq15}
\eeqarr
where $\gamma (\cdot,\cdot)$ denotes the lower incomplete Gamma function, and,
\beqarr
&& \! \! \! \! \! \! \! \! \! \! \! \! \! \! \!
\mu_{x_{m,k}} \! = \!
- \frac{\mu_f B_m \left(\sqrt{E_m}-\sqrt{E_k} \right)}
{\left(B_m - B_k\right)},
\sigma_{x_{m}}^2 \!=\! \frac{B_m}{2},
\sigma_{y_{m}}^2 \!=\! \frac{D_m}{2}, \nn \\
&& \! \! \! \! \! \! \! \! \! \! \! \! \! \! \!
\eta_{m,k} = \left(B_m^{-1} - B_k^{-1}\right), \quad \delta_{m,k} = \left(D_m^{-1} - D_k^{-1}\right),
\eeqarr
using which $\zeta_{m,k}$ and $\theta_{m,k}$ can be computed from (\ref{eq13}) and $\mathcal{K}_{m,k}$ is then found as
\beqarr
\mathcal{K}_{m,k} \! = \! \frac{1}{\theta_{m,k}}
\! \left( \! \frac{1}{2} \ln \left(\frac{B_kD_k}{B_mD_m} \right)
\! + \! \frac{\mu_f^2\left(\sqrt{E_k} - \sqrt{E_m}\right)^2}
{\left(B_k - B_m\right)} \! \right) .
\label{eq16}
\eeqarr
\section{Optimal Multi-level ASK Constellation}
This section presents the optimization framework to minimize the system's SEP and proposes the algorithms to obtain the optimal multi-level ASK constellation to achieve the minimum SEP under the constraint of the availability of a fixed average energy. From the statistics of the channels and the noise, the average SNR per symbol is defined as
\beq
\Gamma_{av} \dn \frac{1}{M}\sum_{m = 1}^{M} \Gamma_m
= \frac{\left(\alpha^2 + \beta\right)\sigma_h^4
+ \sigma_{h_d}^2}{M\sigma_n^2}
\sum_{m=1}^M E_m \, .
\label{eq17}
\eeq
Thus, the optimization problem can be formulated as
\beqarr
&&\!\!\!\!\!\!\!\!\!\!\!\!\!\!\!\!\!\!\!\!\!\!\!\! 
\min_{E_1, E_2, \dots, E_M} P_e \nn \\
&&\!\!\!\!\!\!\!\!\!\!\!\!\!\!\!\!\!\!\!\!\!\!\!\! 
\text{subject to } \frac{1}{M}\sum_{m = 1}^{M}E_m =  \frac{\sigma_n^2\Gamma_{av}}{\left(\alpha^2 + \beta\right)\sigma_h^4 + \sigma_{h_d}^2}.
\label{eq18}
\eeqarr 

From the one-dimensional structure of the ASK constellation and the expression in (\ref{eq15}), it is observed that the SEP is inherently dependent on the inter-symbol separations, and the PEP for the $m$-th symbol is maximally contributed by its neighboring constellation $\left(m-1\right)$-th and $\left(m+1\right)$-th symbols. Thus, the optimization problem in (\ref{eq18}) can be equivalently re-formulated as
\beqarr
&&\!\!\!\!\!\!\!\!\!\!\!\!\!\!\!\!\!\!\!\!\!\!\!\! 
\min_{E_m, \forall m \in \left\{ 1,\ldots,M \right\}} \max\left(P_{e_{m\rightarrow m - 1}}, P_{e_{m\rightarrow m + 1}}\right) \nn \\
&&\!\!\!\!\!\!\!\!\!\!\!\!\!\!\!\!\!\!\!\!\!\!\!\!
\text{subject to} \quad \frac{1}{M}\sum_{m = 1}^{M}E_m =  \frac{\sigma_n^2\Gamma_{av}}{\left(\alpha^2 + \beta\right)\sigma_h^4 + \sigma_{h_d}^2}, \nn \\
&& \quad
P_{e_{1\rightarrow0}} = P_{e_{M\rightarrow M+1}} = 0 \, .
\label{eq19}
\eeqarr

From the receiver structure in \eqref{eq7}, it is evident that the variance of $r \big |s_m$ is directly proportional to $E_m$. Thus, a higher separation between the constellation points would be required for higher values of $E_m$ to achieve a lower PEP value. This implies that for a better SEP performance, we should ensure $d_{m, m-1} < d_{m, m+1}$ for $m = 2,\dots,M-1$, where $d_{m,n}$ denotes the separation between $E_m$ and $E_n$. Furthermore, considering the symmetric nature of the distribution of $r \big| s_m$ about the mean constellation point and $d_{m, m-1} < d_{m, m+1}$, it can be concluded that $P_{e_{m\rightarrow m-1}} > P_{e_{m\rightarrow m+1}}$. Therefore, the optimization problem in \eqref{eq19} can be further simplified as
\beqarr
\label{eq40}
&&\!\!\!\!\!\!\!\!\!\!\!\!\!\!\!\!\!\!\!\!\!\!\!\! 
\min_{E_m , \forall m \in \left\{ 1, \ldots, M \right\}}
P_{e_{m\rightarrow m - 1}}  \nn \\
&&\!\!\!\!\!\!\!\!\!\!\!\!\!\!\!\!\!\!\!\!\!\!\!\!
\text{subject to } \frac{1}{M}\sum_{m = 1}^{M}E_m =  \frac{\sigma_n^2\Gamma_{av}}{\left(\alpha^2 + \beta\right)\sigma_h^4 + \sigma_{h_d}^2}, \nn \\
&&
P_{e_{1\rightarrow0}}= 0 \, .
\eeqarr
In order to solve the optimization problem given in \eqref{eq40}, we first set the initial constellation point to be as low as possible, i.e., $E_1 = 0$. Then, we assume that there exists a constellation with $t^* = P_{e_{m \rightarrow m - 1}}$ which follows the average energy constraint. We find the value of $E_2$ such that $P_{e_{2 \rightarrow 1}} = t^*$. Using this $E_2$, we find the next point $E_3$ such that $P_{e_{3 \rightarrow 2}} = t^*$. This is carried out iteratively to compute all the optimal constellation points. The steps are summarized in Algorithm \ref{alg1}.
\begin{algorithm}[!t]
\caption{Constellation Design Algorithm}\label{alg1}
\begin{algorithmic}[1]
\Function{ConstellationDesign}{$t$}
    \State Initialize $E_1 \gets 0$, $P_{e_{1 \rightarrow 0}} \gets 0$.
    \For{$m = 2$ to $M$}
        \State Find $E_m > E_{m-1}$
        \Statex \hspace{0.93cm} such that for $\mathcal{K}_{m, m-1}$ computed using \eqref{eq16}
        \Statex \hspace{0.93cm} we have $P_{e_{m\rightarrow m-1}} = t$.
    \EndFor
    \State \Return $\left\{E_m\right\}_{m=1}^{M}$
\EndFunction
\end{algorithmic}
\end{algorithm}
\begin{algorithm}[!t]
\caption{Bisection Algorithm}\label{alg2}
\begin{algorithmic}[1]
\Function{Bisection}{}
    \State Initialize $t_{\text{low}} \gets 0$, $t_{\text{high}} \gets 1$.
    \State Set $\mathcal{C} \gets \frac{\sigma_n^2\Gamma_{av}}{\left(\alpha^2 + \beta\right)\sigma_h^4 + \sigma_{h_d}^2}$.
    \Repeat
        \State Compute $t \gets \dfrac{t_{\text{low}} + t_{\text{high}}}{2}$.
        \State $\left\{E_m\right\} _{m = 1}^{M}\gets \Call{ConstellationDesign}{t}$.
        \State Compute $S \gets \dfrac{1}{M} \sum_{m=1}^M E_m$.
        \If{$S < \mathcal{C}$}
            \State Update $t_{\text{low}} \gets t$.
        \Else
            \State Update $t_{\text{high}} \gets t$.
        \EndIf
    \Until{$|t_{\text{high}} - t_{\text{low}}| < \epsilon$ \textbf{and} $|S - \mathcal{C}| < \epsilon$}
    
    \State \Return $\left\{E_m\right\} _{m = 1}^{M}$
\EndFunction
\end{algorithmic}
\end{algorithm}
In this algorithm, the optimal value of $t^*$ can be computed by obtaining the lowest value of $t^* \in \left( 0,1\right)$, which results in an optimal constellation following the given transmission energy constraint. This is efficiently implemented by using the bisection algorithm presented in Algorithm \ref{alg2}, which uses the fact that the minimum value of the average energy per constellation point increases with a decrease in the value of $t^*$ and then utilizes a binary search algorithm to find the optimal $t^*$. Furthermore, we consider a very low value of $\epsilon$, which is set to $10^{-15}$ in Algorithm \ref{alg2}.
\section{Numerical Results and Discussions}
The analytical and optimization frameworks for the RIS-assisted noncoherent wireless system are corroborated via numerical results in this section. 

\begin{figure}[!t]
\centering
\includegraphics[height=2.0in,width=3.4in]{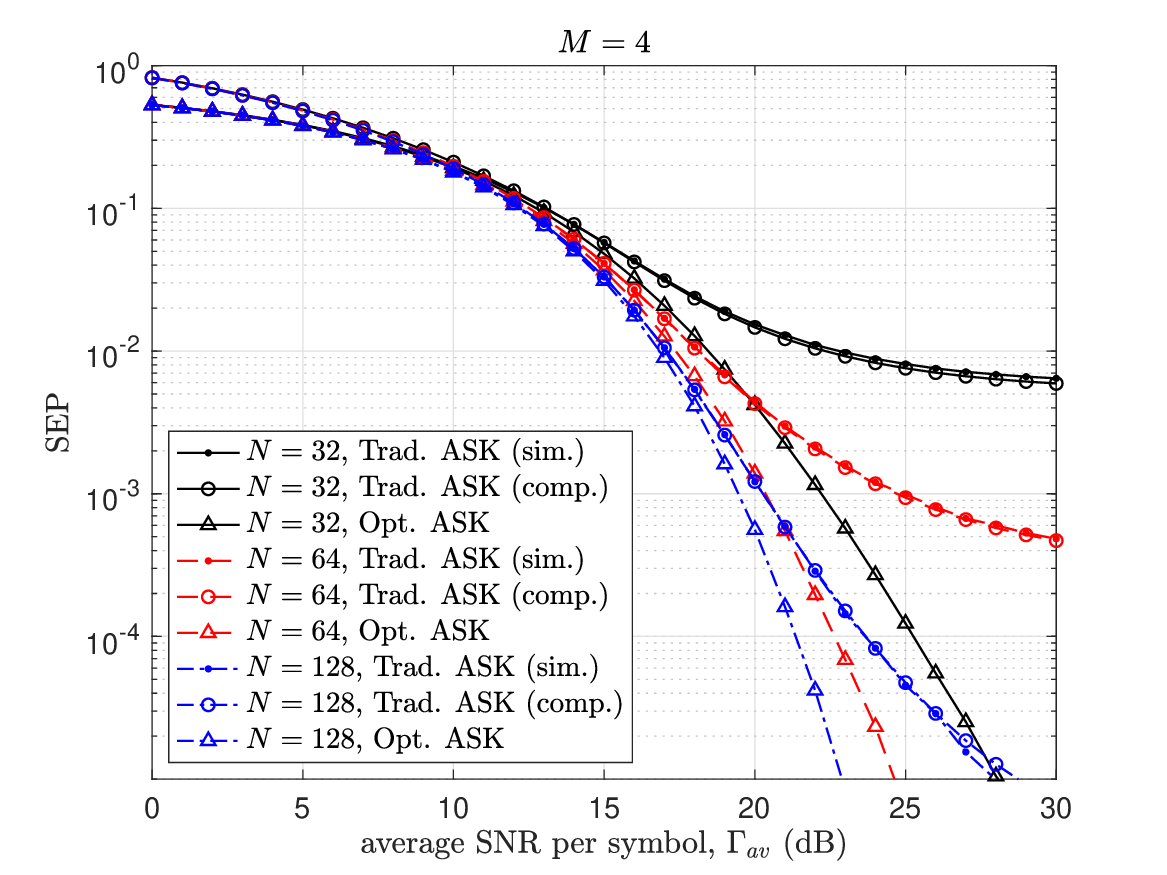}
\caption{SEP versus $\Gamma_{av}$ using traditional and optimal ASK constellations for $M=4$ and $N=32,64,128$.}
\label{f1}
\end{figure}
Fig. \ref{f1} presents the plots of the SEP versus the average SNR per symbol using the traditional equispaced and the optimal 4-level ASK constellations for varying numbers of RIS elements. The simulation plots using the traditional ASK constellation, denoted by `Trad. ASK (sim.)', are generated using Monte Carlo simulations with the receiver structure in (\ref{eq7}) and the computational plots, denoted by `Trad. ASK (comp.)', are obtained by using the traditional ASK constellation points in (\ref{eq15}). The tightness of these plots justifies the correctness of the analytical framework and the validity of the Gamma approximation to be statistically equivalent to the weighted sum in (\ref{eq12}). Furthermore, the plots obtained using the optimal ASK constellation are also presented, and it is observed that the optimal ASK outperforms the traditional ASK in terms of the SEP performance and the diversity order of the system. Moreover, the performance improves with the increase in the number of RIS elements.

\begin{figure}[!t]
\centering
\includegraphics[height=2.1in,width=3.4in]{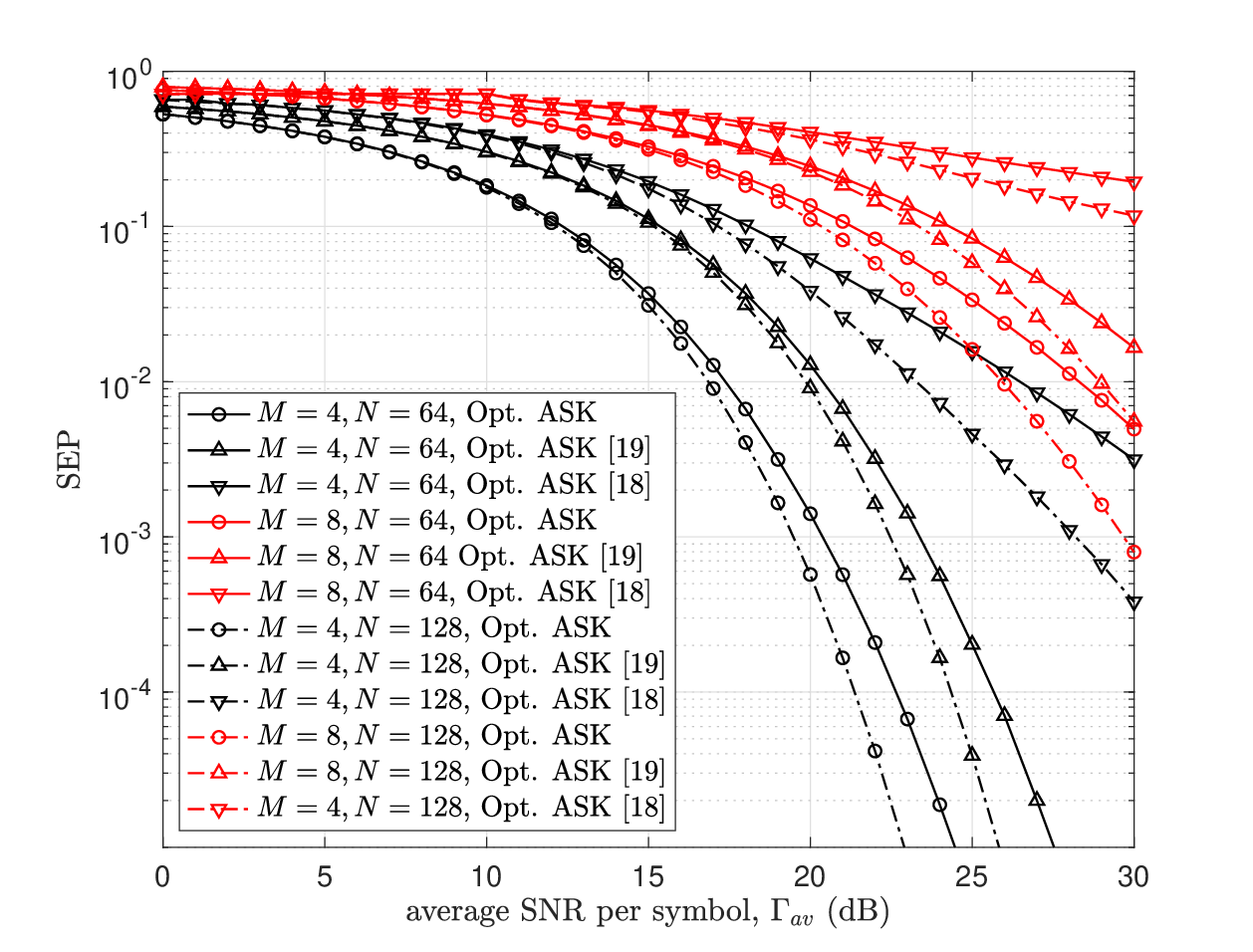}
\caption{SEP versus $\Gamma_{av}$ using optimal ASK constellations and the receiver structures in (\ref{eq7}), \cite{mishra2024optimal}, and \cite{mukhopadhyay2024optimal} for $M=4,8$ and $N=64, 128$.}
\label{f2}
\end{figure}
Fig. \ref{f2} presents the comparison of the plots of the SEP using optimal 4-level and 8-level ASK constellations corresponding to the receiver structure in (\ref{eq7}) and the receiver structures proposed in \cite{mishra2024optimal} and \cite{mukhopadhyay2024optimal}. It is to be noted that the receiver structure in (\ref{eq7}) is optimal as compared to the one in \cite{mishra2024optimal} and \cite{mukhopadhyay2024optimal}, which carry out symbol detection using the energy and the real part of the received symbol, respectively. It is observed that although the diversity order using both the receiver structures is the same, the optimal receiver outperforms the receiver in \cite{mukhopadhyay2024optimal}. Further, the optimal receiver achieves a lower SEP and a higher diversity order compared to the receiver in \cite{mishra2024optimal}. It is also observed that the SEP performance degrades with an increase in the modulation order of the ASK constellation.

\begin{figure}[!t]
\centering
\includegraphics[height=2.1in,width=3.4in]{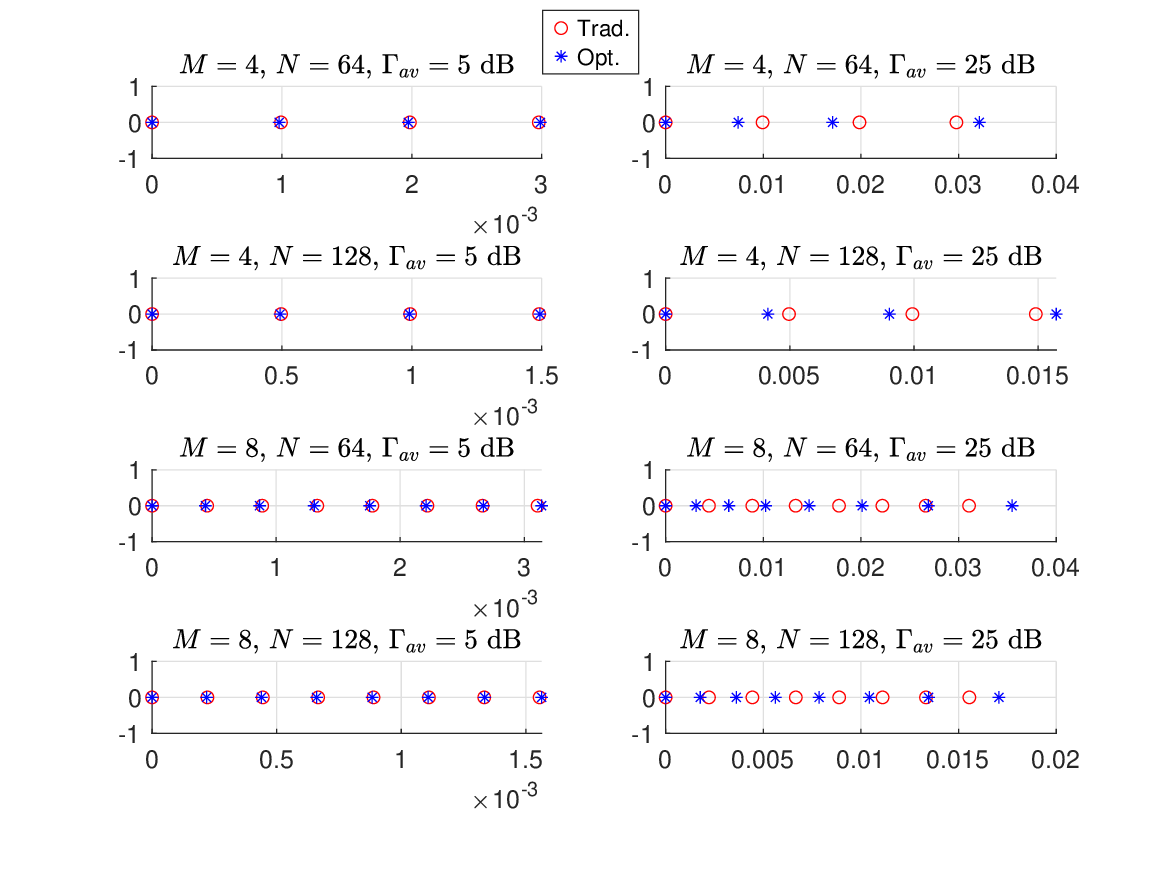}
\caption{Comparison of traditional and optimal one-sided ASK constellations for $M=4,8$, $N=64, 128$, and $\Gamma_{av}=5,25$ dB.}
\label{f3}
\end{figure}
Finally, Fig. \ref{f3} shows the comparison between the constellation plots of the traditional and optimal 4-level and 8-level ASK constellations with varying values of $N$ and SNR. It is observed that the difference between the optimal and the traditional ASK constellations becomes prominent at higher SNR values and lower values of the RIS's reflecting elements. Moreover, it is also observed that the separation between consecutive optimal constellation points increases with increasing value of $E_m$.
\section{Conclusion}
A wireless system was considered in this paper, where the communication between the transceiver pair was assisted by a RIS with $L$ reflecting elements. The transmitter employed multi-level one-sided ASK modulation for data transmission, which was decoded using an optimal noncoherent detection rule at the receiver. A novel statistical framework was proposed to approximate the weighted sum of a non-central and a central chi-squared random variable, which was utilized to derive the closed-form expression for the union bound on the SEP of the system. An optimization framework was proposed to obtain optimal ASK constellations minimizing the SEP under transmission energy constraints, and novel algorithms were presented to obtain the optimal ASKs. Numerical results were presented, and it was observed that the proposed statistical approximation led to tight approximations with the exact SEP results. Furthermore, the superiority of the optimal ASKs as compared to traditional equispaced ASKs was demonstrated in terms of achieving lower SEP values and higher diversity orders. Future directions aim to extend this analysis for RIS-assisted noncoherent MIMO systems.




\bibliographystyle{IEEEtran}
\bibliography{IEEEabrv,references}







\end{document}